\documentclass{aa}
\usepackage{graphicx}
\usepackage{natbib}  
\usepackage{txfonts}
\begin{document}

   \title{GIANO-TNG spectroscopy of red supergiants in the young star cluster RSGC3}
%   \subtitle{}

   \author{L. Origlia\inst{1}
          \and E. Oliva \inst{2}   % {2} = Arcetri
          \and N. Sanna \inst{2} % {2} = Arcetri             
          \and A. Mucciarelli \inst{3} % {3} = Bologna University
          \and E. Dalessandro \inst{3} % {3} = Bologna University
          \and S. Scuderi\inst{4}
          \and C. Baffa \inst{2}
          \and V. Biliotti \inst{2}
          \and L. Carbonaro \inst{2}
          \and G. Falcini  \inst{2}
          \and E. Giani\inst{2}  
          \and M. Iuzzolino\inst{2}  
          \and F. Massi\inst{2}
          \and M. Sozzi\inst{2}
          \and A. Tozzi\inst{2}
          \and A. Ghedina\inst{5}  %TNG
          \and F. Ghinassi\inst{5}  %TNG
          \and M. Lodi\inst{5}
          \and A. Harutyunyan\inst{5}
          \and M. Pedani\inst{5}
          }

\institute{
             INAF - Osservatorio Astronomico di Bologna,
             Via Ranzani 1, I-40127 Bologna, Italy
             \email{livia.origlia@oabo.inaf.it}
         \and
              INAF - Osservatorio Astrofisico di Arcetri,
              Largo E. Fermi 5, I-50125, Firenze, Italy
         \and
             University of Bologna, Physics \& Astronomy Dept.,
             Viale Berti Pichat 6-2, I-40127 Bologna, Italy
         \and
             INAF - Osservatorio Astrofisico di Catania,
              via S. Sofia 78, I-95123 Catania, Italy
         \and
             INAF - TNG, ORM Astronomical Observatory,
              E-38787 Garafia, TF, Spain
             }

\authorrunning{Origlia et al.}
\titlerunning{GIANO-TNG spectroscopy of red supergiants in RSGC3}

   \date{Received .... ; accepted ...}

% \abstract{}{}{}{}{} 
% 5 {} token are mandatory
 
  \abstract
  % context heading (optional), leave it empty if necessary  
   {}
   {The Scutum complex in the inner disk of the Galaxy has a number of young star clusters dominated by red supergiants that are 
heavily obscured by dust extinction and observable only at infrared wavelengths. 
These clusters are important tracers of the recent star formation and chemical enrichment history 
in the inner Galaxy.
   }
   {During the technical commissioning and as a first science verification 
of the GIANO spectrograph at the Telescopio Nazionale Galileo, 
we secured high-resolution (R$\simeq$50,000) near-infrared spectra 
of five red supergiants in the young Scutum cluster RSGC3. 
   }
   {Taking advantage of the full YJHK spectral coverage of GIANO in a single exposure, we were able to 
measure  
several tens of atomic and molecular lines that were suitable for determining chemical abundances.
By means of spectral synthesis and line equivalent width measurements,  
we  obtained abundances of Fe and iron-peak elements such as Ni, Cr, and Cu,
alpha (O, Mg, Si, Ca, Ti), other light elements (C, N, F, Na, Al, and Sc), and some s-process elements (Y, Sr).
We found average half-solar iron abundances and solar-scaled [X/Fe] abundance patterns 
for most of the elements, consistent with a thin-disk chemistry.
We found depletion of [C/Fe] and enhancement of [N/Fe], consistent with standard CN burning, and 
low $\rm ^{12}C/^{13}C$ abundance ratios (between 9 and 11), 
which require extra-mixing processes in the stellar 
interiors during the post-main sequence evolution.   
We also found local standard of rest V$_{\rm LSR}$=106 km/s and heliocentric V$_{\rm hel}$=90 km/s radial velocities
with a dispersion of 2.3 km/s.
   }
   {The inferred radial velocities, abundances, and abundance patterns of RSGC3 are very similar to those previously measured
in the other two young clusters of the Scutum complex, RSGC1 and RSGC2, suggesting a common kinematics and chemistry within the Scutum complex.}

   \keywords{Techniques: spectroscopic --
             stars: supergiants --
             stars: abundances --    
             infrared: stars}

   \maketitle
%
%________________________________________________________________

\section{Introduction}

After RSGC1 \citep{figer06} and RSGC2 \citep{davies07}, 
RSGC3 is the third young star cluster,
to be rich in red supergiant (RSG) stars and located 
at the base of the Scutum-Crux arm 
\citep{clark09,alexander09}, 
at a distance of $\sim$3.5 kpc from the Galactic center.

These clusters host some tens of RSGs (with stellar masses 
of $\sim$14-20 $M_{\odot}$ and ages between 12 and 20 Myr) and 
have an estimated total 
mass of 2-4$\rm \times 10^{4}~M_{\odot}$. 
Some extended associations of RSGs  have been also identified around them: 
Alicante 8, in the proximity of RSGC1 \citep{negueruela10}; 
Alicante 7 \citep{negueruela11}; and Alicante 10 \citep{gonzalez12}, 
have been associated with RSGC3.
\citet{davies07} discussed the possibility that the Scutum 
complex is a giant star-forming region, where the activity 
is triggered by interactions with the Galactic bar, which is believed to 
end close to the base of the Scutum-Crux arm. 
Altogether, the Scutum complex provides a significant proportion of the known 
RSG population in the Galaxy. 

The spectroscopic characterization of the chemical and kinematic properties 
of this region is still largely incomplete,
given the prohibitive  
extinction ($A_{V}>$10 mag) that affects 
the Galactic plane at optical wavelengths, 
and observations can be performed mainly in the IR spectral domain. 

\begin{table*}
\caption{RSG stars in the star cluster RSGC3, observed with GIANO.}
\label{tab1}     
\begin{center}
\begin{tabular}{lllllllllll}  
\hline\hline
Ref & RA (J2000) & DEC (J2000) & SpT & J & H & K & T$_{eff}$ & A$\rm _K$ & RV$_{hel}$ & RV$_{LRS}$\\
\hline
S2 & 18~45~26.54  &–03~23~35.3  &  M3Ia & 8.53  & 6.62  & 5.75  & 3605 & 1.20 & 90 & 106\\ 
S3 & 18~45~24.34  &–03~22~42.1  &  M4Ia & 8.54  & 6.43  & 5.35  & 3535 & 1.47 & 86 & 102\\ 
S4 & 18~45~25.31  &–03~23~01.1  &  M3Ia & 8.42  & 6.39  & 5.31  & 3605 & 1.44 & 90 & 106\\ 
S5 & 18~45~23.26  &–03~23~44.1  &  M2Ia & 8.51  & 6.52  & 5.52  & 3660 & 1.43 & 92 & 108\\ 
S7 & 18~45~24.18  &–03~23~47.3  &  M0Ia & 9.12  & 6.97  & 6.20  & 3790 & 1.40 & 91 & 107\\ 
\hline
\end{tabular}
\end{center}

{\bf Note}: Identification names, coordinates, spectral types,  magnitudes, effective temperatures, 
reddening from \citet{davies09b}; radial velocities (in the heliocentric and local standard of rest reference systems) 
in units of km/s from the GIANO spectra.\\
\end{table*}

First radial velocity measurements  have suggested 
that these young clusters and associations may share a common 
kinematics. For example, \citet{davies07} and \citet{davies08} 
derived average $v_{rad}$=+109 and +123 km/s for RSGC2 and RSGC1, respectively, 
from medium-high resolution K-band spectra of the CO bandheads.
\citet{negueruela11} obtained average +102 and +95 km/s for RSGC2 and RSGC3, respectively,
from their measurement of the Ca~II triplet lines.

Some chemical abundances of Fe, C, O, and other alpha elements have been derived for stars in 
RSGC1 and RSGC2 \citep{davies09b} by using NIRSPEC-Keck spectra at a resolution R$\approx$17,000.
More recently, a few RSGs in RSGC2 and RSGC3 have been observed with GIANO, the new, cross-dispersed NIR spectrograph 
of the Telescopio Nazionale Galileo (TNG)
at the Roque de Los Muchachos Observatory in La Palma (Spain) \citep[see e.g.][]{oli12a,oli12b,oli13,ori14}.
More specifically, out of the targets observed during the commissioning runs to test the science performances of this new 
instrument, our group observed three bright red supergiants (RSGs) in the star cluster RSGC2 in July 2012 and 
five RSGs in the star cluster RSGC3 in October 2013. 
Chemical abundances for the stars in RSGC2 have already been  published in \citet{ori13}. 
The chemical and kinematic analysis of the stars observed in RSGC3 is presented and discussed in this paper.

  \begin{figure*}
  \centering
  \includegraphics[width=\hsize]{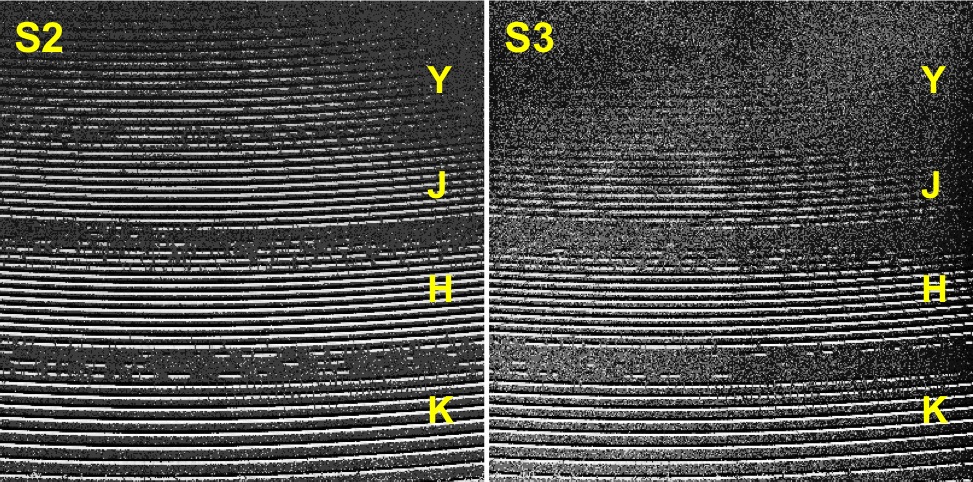}
   \caption{GIANO 2D (A-B) spectra of two observed RSGs in the RSGC3 star cluster.}
              \label{echelle}
    \end{figure*}

\section{Observations and data reduction}

  \begin{figure*}
  \centering
  \includegraphics[width=\hsize]{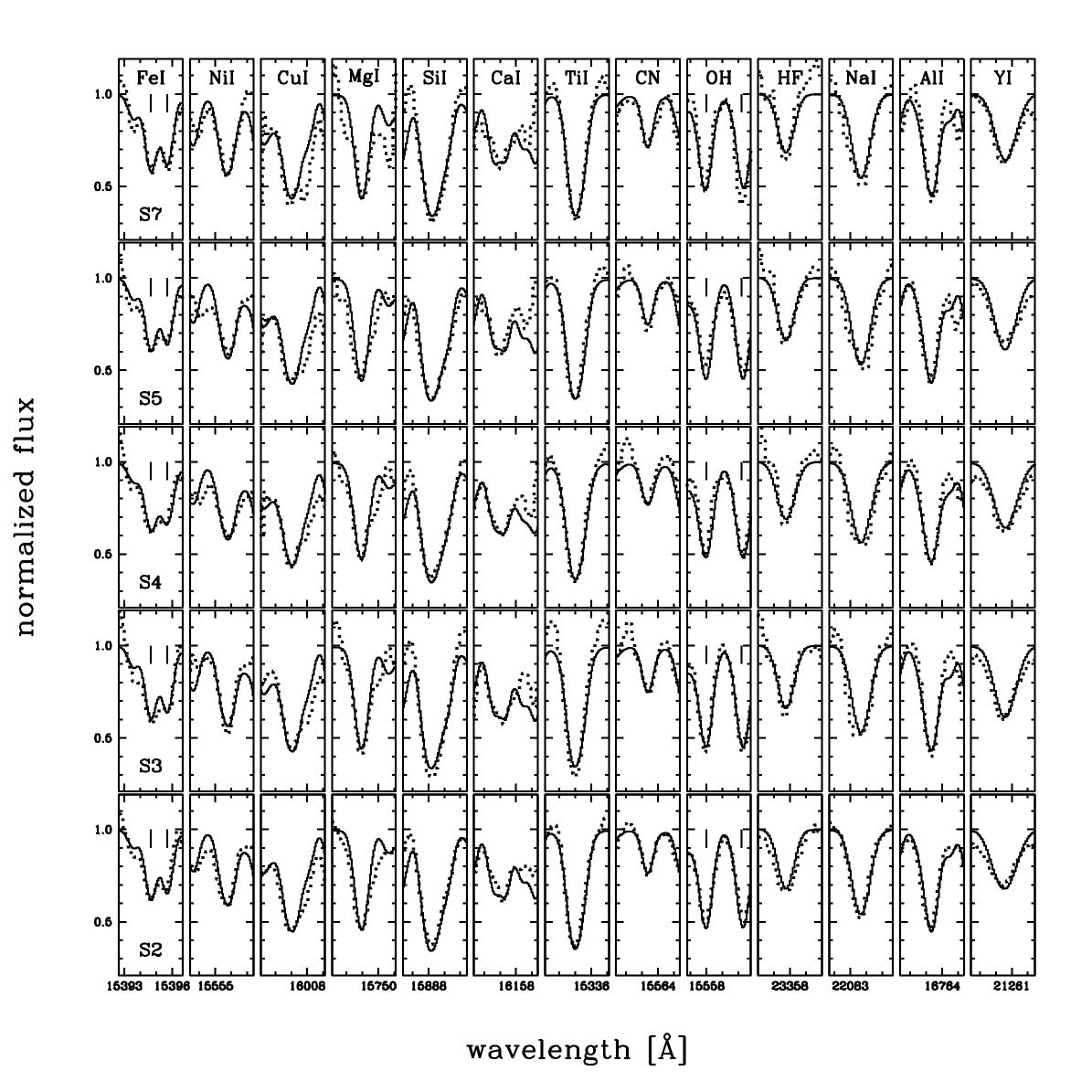}
   \caption{GIANO spectra around  metal lines of interest
            for the five observed RSG stars (dotted lines). Our best-fit models are overplotted as solid lines.
            Rest frame wavelengths are defined in air and thickmarks represent every \AA\ in each panel.}
              \label{figlines}
    \end{figure*}

We used GIANO to observed five stars in RSGC3,  
S2, S3, S4, S5, and S7, from the list of \citet[][see their Tables 1 and 2]{clark09}.
For these RSGs they provided a spectral classification 
and a photometric estimate of the reddening and effective temperature that we list in Table~\ref{tab1}.  
We note that the estimated reddening varies by large amounts (for example up to about two visual magnitudes 
between S2 and S3), even though the stars are clustered within a projected distance
of about 1 arcmin \citep[see e.g., Fig.~2 in][]{clark09}, 
which indicates that the cluster is both heavily and differentially obscured by foreground and/or 
local dust absorption.

The GIANO spectra of the RSGs in RSGC3 were collected during two technical nights in 2013 
October 23 (star S2) and 27 (stars S3, S4, S5, and S7). 
GIANO is interfaced to the telescope with a couple of fibers mounted on the same connector 
at a fixed projected distance on sky of 3 arcsec.
We observed the science targets by nodding on fiber,
i.e. target and sky were taken in pairs and alternatively acquired on Fibers A and B, respectively, 
for an optimal subtraction of the detector noise and background. 
Two pairs of AB observations were performed for a total on-source integration time of 20 min.

An (A-B) 2D-spectrum was computed from each pair of exposures.   
Because of the image slicer, each 2D  frame contains four tracks per order (two per fiber).
Fig.~\ref{echelle} shows the (A-B) 2D echellograms of the S2 and S3 stars as an example.
We note that significant continuum emission in the Y band has been detected only in star S2, 
the least reddened in the observed sample (see Table~\ref{tab1}), while only a marginal detection 
has been obtained in the other stars.

 In a single exposure, the 2D echellogram of GIANO covers
the wavelength range  from 0.95~$\mu$m to 2.4~$\mu$m at a resolving power
R$\simeq$50,000, by spanning 49 orders from \#32 to \#80.
The spectral coverage is complete up to 1.7 $\mu$m.
At longer wavelengths, the orders become larger than the
2k  $\times $ 2k detector. The effective spectral coverage in the K-band is about 75\%.

To extract and perform the wavelength calibration of the echelle orders from the 2D GIANO spectra, 
we  used the ECHELLE package in IRAF and some new, 
ad hoc scripts grouped in a package called GIANO\_TOOLS that can be retrieved from the TNG 
WEB page (http://www.tng.iac.es/instruments/giano/gia\-no\_tools\_v1.2.0.tar.gz). 
We used 2D spectra of a tungsten calibration lamp taken in daytime  to map the geometry of the 
four spectra in each order and for flat-field purposes.
A U-Ne lamp spectrum at the beginning and/or  end of each observing night
was used for wavelength calibration. 
We selected about 30 bright lines (mostly Ne lines) distributed over 
a few orders to obtain a first fit, then the optimal wavelength solution 
was computed by using  about 300 U-Ne lines that were distributed over all orders, with a typical 
accuracy of 300 m/s throughout the entire echellogram.
The positive (A) and negative (B) spectra of the target stars were extracted and added 
together to get a 1D wavelength-calibrated spectrum with the best possible signal-to-noise ratio (S/N).

\section{Spectral analysis and chemical abundances}

From the observed GIANO spectra, we were able to provide radial velocity measurements of the observed RSGs 
(see Table~\ref{tab1}), finding 
average V$_{\rm LSR}$=106 km/s and heliocentric V$_{\rm hel}$=90 km/s values
with a dispersion of 2.3 km/s, which is
fully consistent with the stars being members of the RSGC3 cluster and the Scutum complex 
\citep[see, e.g.,][]{davies07, davies08, negueruela11}.  

The chemical abundances were measured  as in \citet{ori13} for the red supergiants in RSGC2. 
We used an updated version \citep{ori02} of the code,
as first described in \citet{ori93},
to model RSG spectra  by varying the stellar parameters and the
element abundances. 
The code uses the LTE approximation and is based
on both the molecular-blanketed model atmospheres of
\citet{jbk80}, at temperatures $\le $4000~K,
and on the ATLAS9 models for temperatures above 4000~K. 

The NIR  continuum opacity in cool giant and supergiant star atmospheres is mostly set by the H$^-$ ion, 
and therefore it is most important to secure
a correct molecular 
blanketing to define the temperature structure when computing these model atmospheres. 
Metal abundances are much less critical in defining such an opacity for the continuum
 \citep[see also][]{ori05}.

The code also
includes thousands of NIR atomic transitions 
from the Kurucz database\footnote{http://www.cfa.harvard.edu/amp/ampdata/kurucz23/sekur.html},
\citet{bie73}, and \citet{mel99}, while
molecular data are taken from our own \citep[][ and subsequent updates]{ori93} 
and B. Plez (private communications) compilations.
Hyperfine structure splitting has been accounted for in the Ni, Sc, and Cu line profile computations, 
although including them does not significantly affect the abundance estimates.
The reference solar abundances are taken from \citet{gre98}.

As discussed in \citet{ori13}, deriving chemical abundances in RSGs is not trivial.
We used full spectral synthesis 
and equivalent-width measurements of selected lines, 
which are free from significant blending and/or 
contamination by telluric absorption, and which do not have strong wings.
Our best-fit solution is the one that  minimizes the scatter between observed and synthetic spectra,  in terms of both line equivalent width and 
overall spectral synthesis around the lines of interest.

A first compilation of suitable lines at high spectral resolution in the NIR, 
on which to perform abundance analysis in RSGCs, can be found in \citet{ori13}. 
For the RSGs in RSGC3, we were also able to infer some Cu abundances 
from the measurement of the neutral line at $\lambda$=16006.5 \AA\ (log~gf=+0.25, $\chi$=5.35 eV), 
not included in the list by \citet{ori13}.
The presence of possible telluric absorption around each line was carefully 
checked on an almost featureless O-star (Hip103087) spectrum.

The high spectral resolution and the  many CO, OH, and CN molecular lines, as well
as the neutral atomic lines available in the GIANO spectra, allowed us to 
constrain  the atmospheric parameters and 
the spectral broadening quite strictly because of both micro and macro turbulence. 
We obtained a good fit to the observed spectra of the RSGC3 stars by adopting 
{\it i}) 
a stellar temperature T$\approx$3600~K for all the stars except S7, for which we used T$\approx$3800~K, which is in perfect agreement 
with the photometric values reported in Table~\ref{tab1}; {\it ii}) log~g=0.0 for the gravity; {\it iii}) a value of 
3 km/s for the microturbulence velocity; and {\it iv})  
macroturbulence velocity with 
a Gaussian $\sigma $ broadening of about 6  km/s, or the equivalent of a Doppler-broadening 
of 8-9 km/s.  We did not find any other appreciable line broadening by stellar rotation.

Microturbulence velocity in cool giants and supergiants can be constrained at the level of 0.5-1.0 km/s 
(also depending  on the S/N of the spectra) from the shape of the $^{12}$CO bandheads. 
Microturbulence does not change  the depth of the bandheads significantly, it only broadens them. 

As already discussed in \citet{ori13} for the measured RSGs in RSGC2, the observed line profiles in the GIANO spectra 
of the RSGs in RSGC3 are definitely broader than the instrumental one 
(as determined from telluric lines and from the GIANO spectra of standard stars). 
This type of extra-broadening is likely due to macroturbulence and is normally modeled 
in the same way as the instrumental profile by assuming it is Gaussian. 
The inferred velocities are in the range of values (from several to a few tens of km/s) measured, e.g., in the RSGs 
of the Galactic center \citep{ram00,cun07,davies09a}.

At the GIANO spectral resolution of 50,000, we find that variations of $\pm$100K in T$_{eff}$, $\pm$0.5 in log~g and $\pm$0.5 km/s 
in microturbulence have effects on the spectral lines that can also be  noticed  by  visual inspection, and therefore 
we consider them representative of the systematic uncertainties. 
Variations in  temperature, gravity, and microturbulence that are smaller than the above values are difficult to  disentangle 
because of the limited sensitivity of the lines and the degeneracy among stellar parameters. 
Moreover, this kind of  finer tuning would have a negligible 
impact on the inferred abundances (less than a few hundreths of a dex, i.e. smaller than the measurement errors). The impact of using slightly different assumptions for the stellar parameters on the derived 
abundances is discussed in Sect.~\ref{error}.
 
We could estimate the chemical abundances of
Fe, Ni, Cu, Mg, Si, Ca, Ti, Na, Al, Sc, and yttrium  
from the equivalent width measurements of neutral atomic lines in the H and K-bands.
Additional abundances of Cr, Sr, and potassium have only been measured in S2, the least reddenend star, 
by using lines in the Y and J bands.
The Y- and J-band spectra of the other stars do not have enough S/N to obtain 
reliable equivalent width measurements.
Equivalent width measurements of the OH and CN molecular lines in the H-band, and of one HF 
line in the K-band, were used to determine the oxygen,  
nitrogen, and fluorine abundances. 
With regard to the latter, we first estimated the F abundances using the HF(1-0) R9 line parameters from 
\citet{cunha03}, as done in \citet{ori13}.
However,  \citet{jon14} 
 have very recently revised the line parameters 
\citep[see also][]{jor92},
and we also computed the F abundances using these new values
The two abundance determinations differ on average by $\sim $0.2~dex, the higher abundances being 
obtained with the old values from \citet{cunha03}. 
Figure \ref{figlines} shows examples of the spectral lines observed with GIANO 
from which metals abundances have been derived.

\begin{figure}
  \centering
  \includegraphics[width=\hsize]{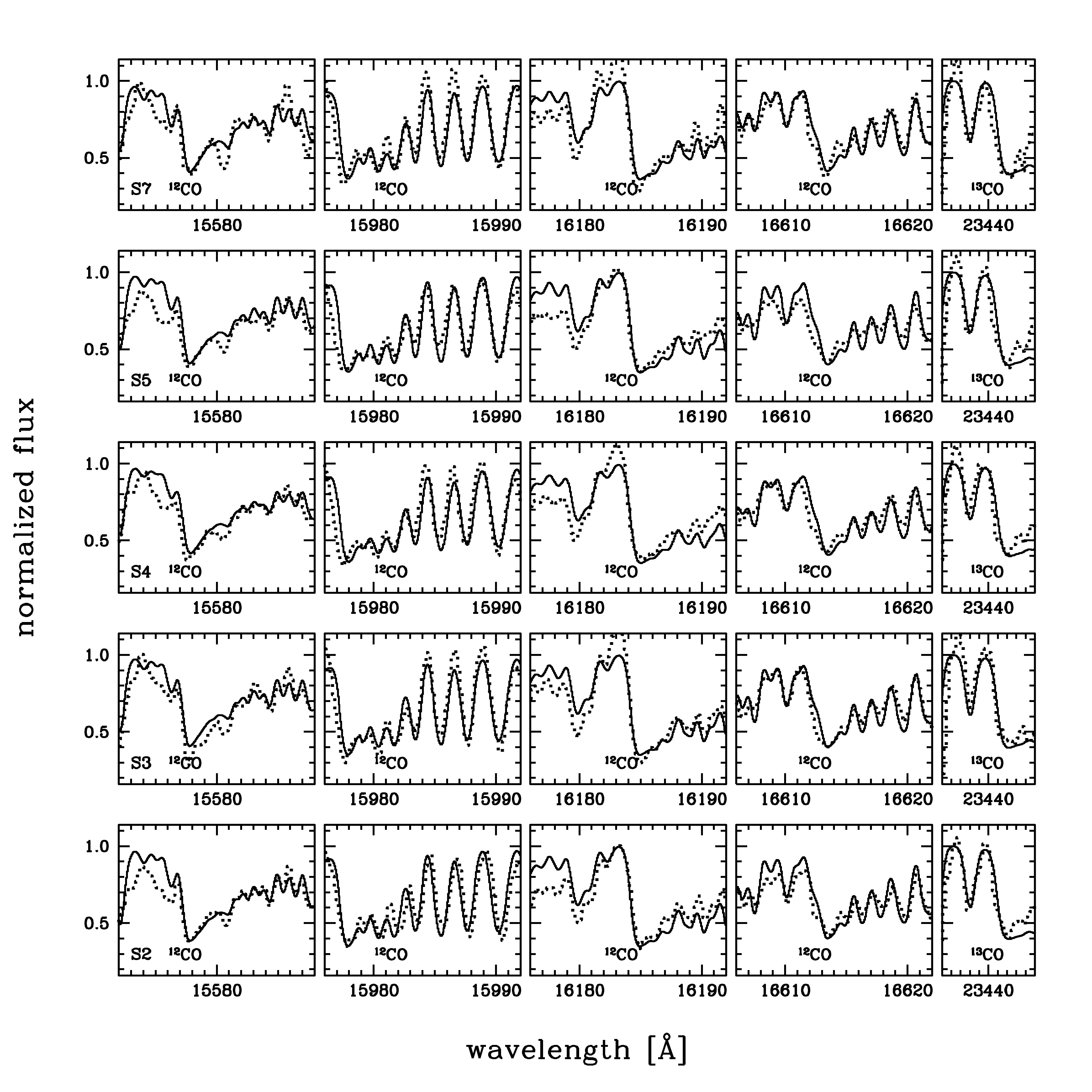}
   \caption{GIANO H-band spectra of the $^{12}$CO (3-0), (5-2), (6-3) and (8-5) bandheads and 
K-band spectra of the $^{13}$CO (2-0) bandhead  
            for the five observed RSG stars (dotted lines). Our best-fit models are overplotted as solid lines.
	    Rest frame wavelengths are defined in air and thickmarks represent every \AA\ in each panel.}
              \label{figco}
    \end{figure}

\begin{figure}
   \centering
  \includegraphics[width=\hsize]{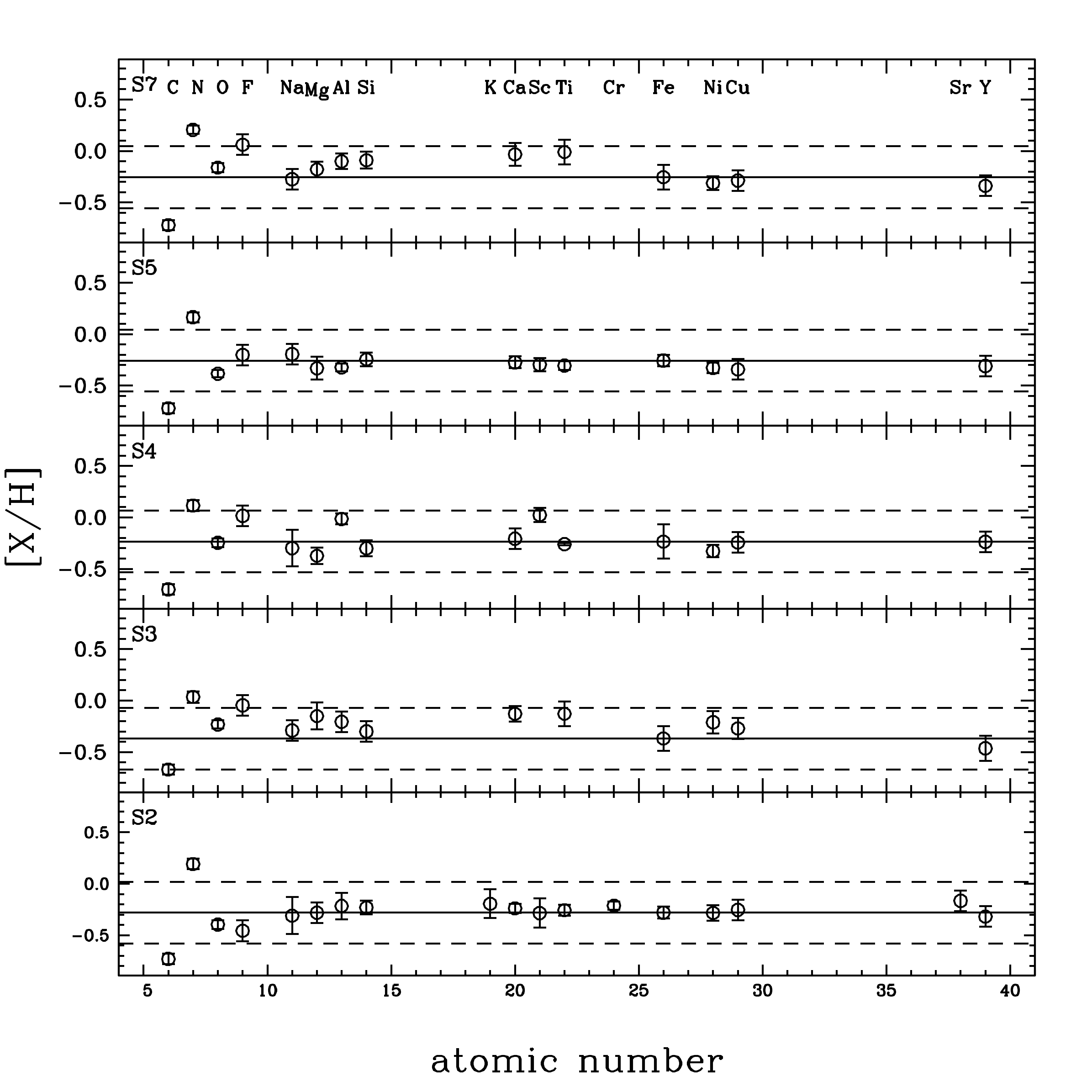}
   \caption{[X/H] chemical abundances as a function of the atomic number for the five RSGs in the star cluster RSGC3,
as observed with GIANO.  The horizontal lines indicate the [Fe/H] abundance (solid) and the $\pm$0.3 dex values (dashed),
as inferred in the present analysis.}
              \label{figabun}
    \end{figure}

The
$^{12}$C and $^{13}$C carbon abundances were mostly determined from the CO bandheads in the 
H- and K-bands, respectively, using full spectral synthesis because of the high level of 
crowding and blending of the CO lines in these stars.
Figure~\ref{figco} shows the GIANO spectra of the five RSGs centered on some of the CO bandheads used 
to determine the abundance.
The average metal abundances for the sampled chemical species are quoted in Table~\ref{tab2} 
and plotted in Fig.~\ref{figabun}.

\subsection{Error budget}
\label{error}

The typical random error of the measured line equivalent widths is between 10 and 20 m\AA, mostly 
the result of a $\pm $2\% uncertainty in the placement of the pseudocontinuum, as estimated 
by overlapping the synthetic and the observed spectra. 
This error corresponds to abundance variations 
ranging from a few hundredths to one-tenth of a dex and 
is lower than the typical 1$\sigma $ scatter  ($\le$0.15 dex) in the derived abundances 
from different lines. 
The errors for the final abundances quoted in Table~\ref{tab2} were obtained 
by dividing these 1$\sigma $ errors by the squared root of the number of used 
lines, typically a few in the case of atomic lines, and 10-20 in the case of the CN and OH 
molecular lines.  When only one line was available, we assumed a 0.1~dex error.

As detailed in \citet{ori13}, a somewhat conservative estimate of the overall systematic uncertainty in the abundance 
(X) determination, which is caused by 
variations of the atmospheric parameters, can be computed with the following formula:
$\rm (\Delta X)^2 = (\partial X/\partial T)^2 (\Delta T)^2 + (\partial X/\partial log~g)^2 (\Delta log~g)^2  + (\partial X/\partial \xi)^2 (\Delta \xi)^2$.
We computed test models with variations of  
$\pm $100~K in temperature, $\pm$0.5~dex in log~g, and $\pm $0.5 km/s in microturbulence velocity
with respect to the best-fit parameters, and we found that these systematic uncertainties in stellar parameters 
can impact the overall abundances at a level of 0.15-0.20 dex. 

As in \citet{ori13}, we 
also determined the statistical significance of our best-fit solution for the spectral synthesis of the 
CO features and the derived carbon abundances.
As a figure of merit of the statistical test, we adopted
the difference between the model and the observed spectrum.
To quantify systematic discrepancies, this parameter is
more powerful than the classical $\chi ^2$ test, which  instead
is equally sensitive to {\em random} and {\em systematic} deviations
\citep[see][for more discussion and references]{ori04}.

Our best-fit solutions always show $>$90\% probability
of being representative of the observed spectra, while
solutions with $\pm$0.1~dex abundance variations or 
with $\rm \Delta T=\pm$100~K, $\rm \Delta log~g=\pm$0.5~dex, and
$\rm \Delta \xi=\pm$0.5~km~s$^{-1}$, as well as corresponding simultaneous variations
in the C abundance (0.1-0.2~dex) to reproduce the depth of the
molecular features, are statistically significant only at 1-2$\sigma$ level. 

As discussed in \citet{dav13} 
\citep[see also][]{lev05} significantly different temperatures can be inferred
using different scales. This issue has been explored by extensive simulations in \citet{ori13} 
for the three studied RSGs in RSGC2, and finding that, 
when temperatures were both significantly warmer and cooler than those that provide the best fit solutions,
one can barely fit the observed spectra except with  very peculiar CNO abundance patterns.
We repeated the same tests for the five RSGs  analyzed here and reached very similar conclusions. 

\begin{table}
\caption{Chemical abundances of the RSG stars in the star cluster RSGC3 observed with GIANO.}
\label{tab2}     
\tabcolsep 3pt
\begin{tabular}{llllll}  
\hline\hline
El$^a$ & \multicolumn{5}{c}{[X/H]$^b$} \\ 
                   & S2 & S3 & S4 & S5 &S7\\
\hline
Fe (26) & -0.28 (12) & -0.37 (5) & -0.23 (4) & -0.26 (11) & -0.25 (4) \\ 
 & $\pm$ 0.06 & $\pm$ 0.12 & $\pm$ 0.17 & $\pm$ 0.06 & $\pm$ 0.12 \\ 
Cr (24) & -0.21 (5) & --   & --         & --         & --               \\ 
 & $\pm$ 0.04 & --          & --          & --          & --     \\ 
Ni (28) & -0.28 (6) & -0.21 (5) & -0.33 (6) & -0.33 (4) & -0.31 (7) \\ 
 & $\pm$ 0.07 & $\pm$ 0.11 & $\pm$ 0.06 & $\pm$ 0.05 & $\pm$ 0.07 \\ 
Cu (29) & -0.25 (1) & -0.27 (1) & -0.24 (1) & -0.34 (1) & -0.29 (1) \\ 
 & $\pm$ 0.10 & $\pm$ 0.10 & $\pm$ 0.10 & $\pm$ 0.10 & $\pm$ 0.10 \\ 
Mg (12) & -0.28 (4) & -0.15 (4) & -0.37 (3) & -0.33 (4) & -0.18 (3) \\ 
 & $\pm$ 0.10 & $\pm$ 0.13 & $\pm$ 0.08 & $\pm$ 0.11 & $\pm$ 0.07 \\ 
Si (14) & -0.23 (5) & -0.30 (1) & -0.30 (3) & -0.25 (3) & -0.09 (2) \\ 
 & $\pm$ 0.07 & $\pm$ 0.10 & $\pm$ 0.08 & $\pm$ 0.07 & $\pm$ 0.08 \\ 
Ca (20) & -0.24 (5) & -0.13 (4) & -0.21 (4) & -0.27 (5) & -0.03 (4) \\ 
 & $\pm$ 0.05 & $\pm$ 0.08 & $\pm$ 0.10 & $\pm$ 0.06 & $\pm$ 0.11 \\ 
Ti (22) & -0.26 (16) & -0.13 (3) & -0.26 (3) & -0.31 (8) & -0.01 (3) \\ 
 & $\pm$ 0.05 & $\pm$ 0.12 & $\pm$ 0.01 & $\pm$ 0.04 & $\pm$ 0.12 \\ 
C  (6) & -0.73       & -0.67      &  -0.70     & -0.72      & -0.62      \\ 
 & $\pm$ 0.05 & $\pm$ 0.05 & $\pm$ 0.05 & $\pm$ 0.05 & $\pm$ 0.05 \\ 
N  (7) &  0.19 (13) &  0.04 (9) &  0.12 (12) &  0.16 (12) &  0.21 (14) \\ 
 & $\pm$ 0.05 & $\pm$ 0.06 & $\pm$ 0.05 & $\pm$ 0.05 & $\pm$ 0.04 \\ 
O  (8) & -0.40 (9) & -0.23 (16) & -0.25 (17) & -0.38 (13) & -0.16 (19) \\ 
 & $\pm$ 0.04 & $\pm$ 0.04 & $\pm$ 0.04 & $\pm$ 0.03 & $\pm$ 0.05 \\ 
F$^c$ (9) & -0.46 (1) & -0.05 (1) &  0.02 (1) & -0.20 (1) &  0.06 (1) \\ 
 & $\pm$ 0.10 & $\pm$ 0.10 & $\pm$ 0.10 & $\pm$ 0.10 & $\pm$ 0.10 \\ 
Na (11) & -0.31 (2) & -0.29 (1) & -0.30 (2) & -0.19 (1) & -0.27 (1) \\ 
 & $\pm$ 0.18 & $\pm$ 0.10 & $\pm$ 0.18 & $\pm$ 0.10 & $\pm$ 0.10 \\ 
Al (13) & -0.22 (3) & -0.21 (1) & -0.01 (3) & -0.32 (3) & -0.10 (2) \\ 
 & $\pm$ 0.13 & $\pm$ 0.10 & $\pm$ 0.05 & $\pm$ 0.04 & $\pm$ 0.08 \\ 
K  (19) & -0.19 (2) & --         & --         & --         & --         \\ 
 & $\pm$ 0.14 & --         & --          & --          & --       \\ 
Sc (21) & -0.28 (3) & --         &  0.02 (2) & -0.30 (2) & --         \\ 
 & $\pm$ 0.14 & --         & $\pm$ 0.07 & $\pm$ 0.06 & --         \\ 
Sr (38) & -0.16 (1) & --         & --         & --         & --         \\ 
 & $\pm$ 0.10 & --         & --         & --         & --         \\ 
Y  (39) & -0.32 (1) & -0.46 (2) & -0.24 (1) & -0.31 (1) & -0.34 (1) \\ 
 & $\pm$ 0.10 & $\pm$ 0.12 & $\pm$ 0.10 & $\pm$ 0.10 & $\pm$ 0.10 \\ 
\hline\hline
$\rm ^{12}C/^{13}C$  & 11$\pm1$ & 11$\pm1$ & 10$\pm1$ & 9$\pm1$ &10$\pm1$ \\
\hline
\end{tabular}

{\bf $^a$} Chemical element and corresponding atomic number in parenthesis.

{\bf $^b$} Numbers in parenthesis refer to the number of lines used to compute the abundances.

{\bf $^c$} F abundances are obtained from the HF(1-0) R9 line using the parameters listed in \citet{jon14}. 

\end{table}

\section{Discussion and conclusions}

Red supergiants are very luminous NIR sources even in 
highly reddened environments, such as the inner Galaxy, and  
they can be spectroscopically studied at high resolution even with 4m-class telescopes,
provided that these are equipped with efficient cross-dispersed spectrographs like GIANO. 
Indeed, the simultaneous access to the YJHK bands provided by GIANO offers the possibility 
of sampling most of the chemical elements of interest and 
to measure from a few to several tens of lines per element, thus enabling an accurate 
and statistically significant abundance analysis.  

All  five RSGs observed in RSGC3 show similar iron abundances, further confirming their likely membership in the cluster.
We found average half-solar iron and iron-peak (Ni and Cu) abundances and solar-scaled [X/Fe] abundance ratios 
(with r.m.s $<0.1$ dex) for the measured s process (yttrium in all the five stars and Sr in only star S2), 
alpha, and other light elements, 
but carbon, nitrogen and, to a lower level, fluorine.
Abundance ratios of [C/Fe] and [N/Fe] are depleted and enhanced, respectively, by a similar factor (between two and three),
and the resulting [(C+N)/Fe] and [(C+N+O)/Fe] average abundance ratios turn out to be about zero with rms of 
0.09 and 0.19 dex, respectively, which is consistent with standard CNO nucleosynthesis.
We also found low  
$\rm ^{12}C/^{13}C$ ratios (between 9 and 11). 
As suggested by \citet{davies09b} and also  mentioned in \citet{ori13} 
rotationally enhanced mixing \citep[e.g.,][]{mey00} can account for both the overall 
depletion of carbon and the low $\rm ^{12}C/^{13}C$ isotopic ratio in RSGs. 
In this respect, it is interesting to note that measurements of the $\rm ^{12}C/^{13}C$ ratio in the interstellar medium \citep{mil05} 
suggest a value of $\sim33$ at a Galactocentric distance of 3.5 kpc.
[F/Fe] and [F/O] are slightly enhanced (by about 0.15 dex), 
which is consistent with  measurements in RSGC2 \citep{ori13} and  
in the Orion Nebula Cluster \citep{cun05}. 

Mild abundance ratio enhancement (at a level of $\sim$0.2 dex) of [Mg/Fe] in star S3, of [Ca/Fe] and [Ti/Fe] 
in stars S3 and S7, and of [Al/Fe] and [Sc/Fe] in star S4 has been measured.
In the least reddened star S2, 
we also measured [K/Fe] and [Sr/Fe] abundance ratios. These are only slightly enhanced (by 0.1 dex), but are still consistent with solar-scaled values 
within the errors.

These abundances and abundance patterns are fully consistent with those measured in RSGC1 and RSGC2 \citep{davies09b,ori13},
and suggest a rather homogeneous kinematics and chemistry within the Scutum complex, as traced by its young populations of RSGs.
Their slightly sub-solar metallicity is intriguing,  as already mentioned in \citet{ori13},
since both measurements of Cepheids 
\citep[see e.g.][ and references therein]{gen13,and13} and open clusters \citep[see e.g.][]{mag09}, as well as model predictions 
\citep[see e.g.][]{ces07}, suggest metal abundances that are well in excess of solar at  larger Galactocentric distances ($\ge$4 kpc).
In this respect, however, it is interesting to note that  the RSGs in the Galactic center \citep[about solar,][]
{ram00,cun07,davies09a,ryd15}
do not also reach these high metal abundances, thus suggesting that the chemical enrichment in the innermost regions of the Galaxy 
is more complicated and deviates from the metallicity gradient observed in the outer disk.
The about solar-scaled metal abundances of alpha and other light elements, similar to those measured in the thin disk 
\citep[see e.g..][]{reddy03}, 
indicate an enrichment by both type II and type I supernovae on long timescales.  

Taken together, these first NIR spectroscopic studies  
at medium-high resolution of the chemical and kinematic properties of the young stellar populations in the inner Galaxy 
 reveal their potential astrophysical impact   
in complementing the information derived from the older and fainter population of giant stars in 
less obscured regions. 
Indeed, a proper characterization of the chemical enrichment of the various stellar populations 
across the entire disk is critical for a comprehensive reconstruction of its formation and evolution history.
 
Finally, it is worth noting that these high-resolution studies of RSGs in the Galaxy and in the Magellanic Clouds 
are also crucial for better interpreting the 
low resolution and/or integrated spectral information of these young stellar populations 
in distant star-forming galaxies. 

\begin{acknowledgements}
We thank the anonymous referee for useful comments and suggestions.
\end{acknowledgements}

\end{document}